\documentclass[aps,prb,twocolumn,groupedaddress,amsmath,amssymb]{revtex4}

\usepackage{amssymb}
\usepackage{amsmath}
\usepackage{graphicx}
\usepackage{subfigure}
\usepackage{textcomp}
\usepackage{color}
\usepackage{amsfonts}
\usepackage{bbold}
\usepackage{dsfont}
\usepackage{epsfig}
\usepackage{hyperref}

\newcommand{\arctanh}[1]{\operatorname{arctan}}

\begin{document}

\title{Electric field control of valence tautomeric interconversion in Cobalt dioxolene}

\author{A. Droghetti and S. Sanvito}
\affiliation{School of Physics and CRANN, Trinity College, Dublin 2, Ireland}

\date{\today}

\begin{abstract}
We demonstrate that the critical temperature for valence tautomeric interconversion in Cobalt  dioxolene complexes can be 
significantly changed when a static electric field is applied to the molecule. This is achieved by effectively manipulating the 
redox potential of the metallic acceptor forming the molecule. Importantly our accurate density functional theory calculations 
demonstrate that already a field of 0.1~V/nm, achievable in Stark spectroscopy experiments, can produce a change in the critical 
temperature for the interconversion of 20~K. Our results indicate a new way for switching on and off the magnetism in a magnetic 
molecule. This offers the unique chance of controlling magnetism at the atomic scale by electrical means.
\end{abstract}

\maketitle


The integration of molecular electronics \cite{Mol} with spintronics \cite{Spin} has given birth to the new and 
fascinating field of {\it molecular spintronics} \cite{Alex,Dediu}. This at present encompasses a variety of 
phenomena, ranging from spin-injection and transport in organic semiconductors \cite{Dediu} to the 
engineering of magnetic surfaces by means of organic molecules \cite{Spinterface}. A fascinating branch 
of molecular spintronics concerns electron transport in single magnetic molecules \cite{VanDerZant}. Here the 
aim is that of investigating the mutual influence between the spin state of the molecule and the electrical 
current that flows across it. 

The detailed knowledge of the $I$-$V$ curve of a two-terminal device may enable one to read the spin state of 
the molecule~\cite{srft1,Das}. Much more challenging is the task of manipulating such a spin state. Magnetic fields 
do not have enough spatial resolution to address selectively single molecules, 
hence the possibility to use either electric fields or currents appears particularly attractive. Recently it was 
proposed that spin crossover can be achieved by electrical means, simply by exploiting either the first \cite{nadjib} 
or the second order \cite{kim} Stark effect. The idea is that the high-spin (HS) and the low-spin (LS) state of a 
molecule in general have different polarizabilities and different permanent electrical dipoles. Thus the energy 
Stark shift depends on the magnetic state and one can speculate that there exists a particular condition where 
a HS to LS crossover is possible. This would have spectacular consequences on the molecule $I$-$V$ 
characteristic \cite{Sujeet}.

Intriguingly there is an entire class of magnetic molecules, mainly incorporating Fe(II), known to exhibit 
temperature-induced LS to HS spin-crossover (SC) \cite{SC}. Such entropy driven transition is 
regulated by the relative Gibbs free energy between the HS and LS states,
\begin{equation}
\Delta G=G_\mathrm{HS}-G_\mathrm{LS}=\Delta H-T \Delta S\label{deltaG}\:,
\end{equation}
where $\Delta H =H_\mathrm{HS}-H_\mathrm{LS}$ and $\Delta S =S_\mathrm{HS}-S_\mathrm{LS}$ are 
respectively the enthalpy and the entropy variation ($\Delta G>0$ means that the thermodynamically stable 
configuration is LS). In general in SC molecules $\Delta H>0$ but $S_\mathrm{HS}>S_\mathrm{LS}$, 
so that for temperatures large enough the entropic term dominates over the enthalpic one and the molecules 
transit from a LS to a HS configuration. The microscopic mechanism for the SC consists in moving an electron 
from a non-bonding to an antibonding state. The transfer produces the breathing of the metal ion coordination 
sphere with consequent phonon modes softening. Thus, both the spin and the vibrational entropy of the HS state 
dominate over those of the LS ones. Interestingly, it was recently demonstrated that the SC is strongly affected 
by electrostatic perturbations in the crystal environment \cite{Robert}.

\begin{figure}[ht]
\centering\includegraphics[scale=0.35,clip=true]{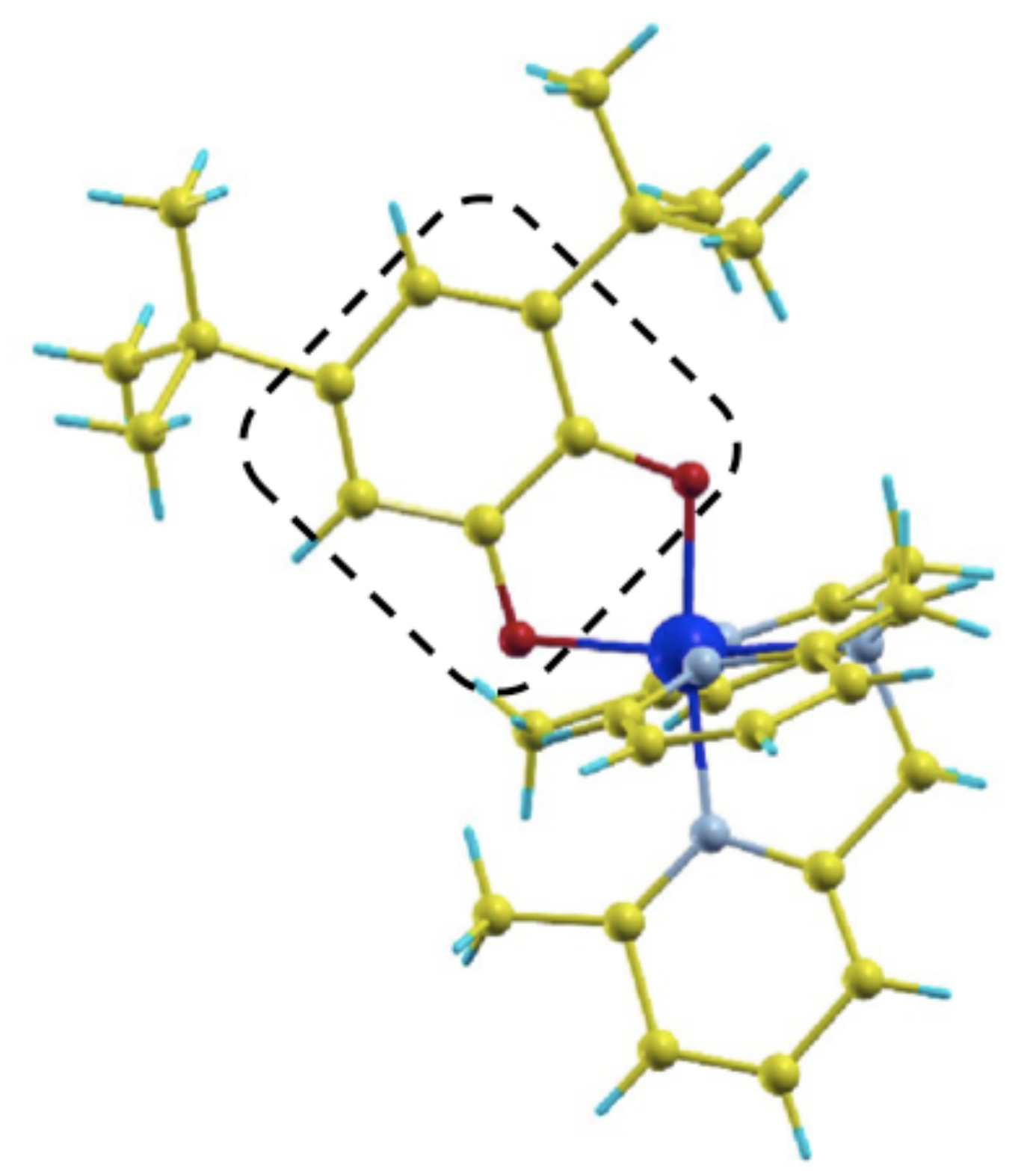}
\caption{The Co-dioxolene complex investigated in this work. The entire cationic unit of the complex Me$_2$tpa=methil 
derivatives of tris(2-pyridylmethyl)amine, DBCat $=3,5$-di-$tert$-butylcatecolato is presented and the o-dioxolene is 
enclosed in the dashed box. Color code: C=yellow, O=red, Co=grey (large sphere), N=grey (small sphere), H=blue.}
\label{Fig1}
\end{figure}
An effect closely resembling the SC is observed in Co-dioxolene complexes, which exhibit an interconversion 
between redox isomers \cite{Hendrickson,Dei}. This is called valence tautomeric interconversion (VTI). The complex
[Co(Me$_2$tpa)(DBCat)](PF$_6)$ (Me$_2$tpa=methil derivatives of tris(2-pyridylmethyl)amine, 
DBCat $=3,5$-di-$tert$-butylcatecolato)\cite{Giordano,Giordano_2}, is an example of these compounds (see Fig.~\ref{Fig1}). 
There are two key ingredients in such a class of molecules: the transition metal ion and the 
o-dioxolene group. In general, o-dioxolenes bind to metal atoms in three different oxidation states: di-negative catecholate 
(Cat), mono-negative semiquinonate (SQ) and neutral quinone (Q). The low covalent character of the bond between 
o-dioxolene and Co allows charge and spin to be localized so that the molecule maintains well defined oxidation states 
and magnetic moments.  

[Co(Me$_2$tpa)(DBCat)](PF$_6)$ is found in crystals as either diamagnetic Co(III)-catecholate (LS Co$^{3+}$-Cat) at 
low temperature, or as paramegnetic high-spin Co(II)-semiquinonate (HS Co$^{2+}$-SQ) at high-temperature. The transition 
from LS Co$^{3+}$-Cat to HS Co$^{2+}$-SQ is driven by an intramolecular electron transfer from the catecholate to 
the Co ion. Similarly to standard SC also the VTI is entropy driven and can be understood with equation~(\ref{deltaG}), where 
LS and HS now correspond respectively to LS Co$^{3+}$-Cat and HS Co$^{2+}$-SQ. 

Much recent effort has been dedicated to increasing the cooperative nature of the VTI in crystals for data storage applications. 
Co-dioxolene complexes however might also find an interesting place in molecular spintronics. In fact the strong interplay 
between charge transfer and spin transition suggests that electric fields might be able to profoundly affect the molecule 
magnetic properties. This is demonstrated here. Our density functional theory (DFT) calculations show that charge transfer 
with consequent spin crossover can be induced by a static electric field. Although we find that the critical fields for the VTI are 
rather large ($\sim1$~V/\AA), we also show that much more modest fields ($\sim0.05$~V/\AA) are enough for changing the 
VTI critical temperature by as much as 100~K. This intriguingly new effect, which may allow one to fine tune electrostatically 
the magnetic state of a molecule, can be readily verified experimentally.  

Achieving an accurate quantitative description of the Co-dioxolene electronic structure is a difficult task. One in fact has to
simultaneously assign the oxidation states of both the Co and the dioxolene group, the Co spin state and also, in the case of 
HS Co$^{2+}$-SQ, the magnetic exchange coupling between Co and SQ. Local and semi-local approximations to DFT are inadequate 
since they largely overestimate the electronic coupling between the Co $3d$ shell and the molecular orbitals of both SQ and Cat. 
This is the same shortfall encountered when describing $F$ centers in wide-gap semiconductors and it is rooted in 
the self-interaction error \cite{Andrea}. In order to overcome this limitation we have employed two approaches. The first is the atomic 
self-interaction correction (ASIC) scheme \cite{ASIC}, so far used for electron transport \cite{Cormac} and for a broad range of 
problems in solids \cite{Alessio}. The second is the standard B3LYP hybrid exchange correlation functional \cite{B3LYP}. 

Calculations are performed with a development version of the SIESTA code~\cite{SIESTA} implementing the ASIC scheme. Norm-conserving 
Troullier-Martin pseudopotentials are employed together with a basis set of double-zeta plus polarization quality. In addiction the GAMESS 
\cite{GAMESS} and the NWCHEM \cite{NWCHEM} quantum chemistry packages are used for the B3LYP calculations. In this case the basis is 
$6$-$31$G$^*$. We consider the experimental molecular geometries determined by standard X-ray data analysis for single crystals 
(Table~\ref{Tab1}) without further geometrical optimization except for the H atoms. This is the best computational strategy since it is at 
present unclear whether DFT has enough accuracy to correctly describe the metal-to-ligand bond lengths across the VTI. Note that the 
same procedure is often adopted when studying SC compounds \cite{Casida}.

\begin{table}[ht!]\centering
\begin{tabular}{lccc}\hline\hline
   Isomer   & $d_{\mathrm{Co-O}}$ (\AA)  & $d_{\mathrm{Co-N}}$ (\AA) & $E$ (eV) \\ \hline\hline
 LS Co$^{3+}$-Cat & 1.883, 1.898 & 2.0, 1.947, 2.011, 2.022 & 0  \\
 HS Co$^{2+}$-SQ & 2.004, 2.083 & 2.169, 2.160, 2.1474, 2.151 & 2.229 \\  
 LS Co$^{2+}$-SQ & 1.883, 1.898 & 2.0, 1.947, 2.011, 2.022  & 2.789 \\  \hline\hline
\end{tabular}
\label{Tab1}
\caption{Co-O ($d_{\mathrm{Co-O}}$) and Co-N ($d_{\mathrm{Co-N}}$) bond lengths for LS Co$^{3+}$-Cat and  HS Co$^{2+}$-SQ as 
determined by X-Ray and used in our calculations. The last column reports the B3LYP total energy of the 
given configuration, calculated with respect to the total energy of LS Co$^{3+}$-Cat. LS Co$^{2+}$-SQ is a fixed spin configuration 
calculated at the molecular geometry of HS Co$^{2+}$-SQ.}
\end{table}

Table \ref{Tab1} displays the energy difference between LS Co$^{3+}$-Cat, LS Co$^{2+}$-SQ and the HS Co$^{2+}$-SQ showing that 
B3LYP qualitatively reproduces the relative order of the states. The absolute values of the relative energies appear however drastically
overestimated with respect to experimentally available  enthalpies (a few hundreds meV) \cite{Bencini, Beni}. Such a quantitative 
disagreement between DFT and experiments, pathological to many SC compounds \cite{Casida,Robert,Reiher}, cannot be attributed 
to having neglected the vibrational zero-point energies, which are much smaller than the electronic ones. At present the reasons
for these quantitative DFT failures are unclear but importantly they affect only marginally our main results. 

\begin{figure}[ht]
\includegraphics[scale=0.25,clip=true]{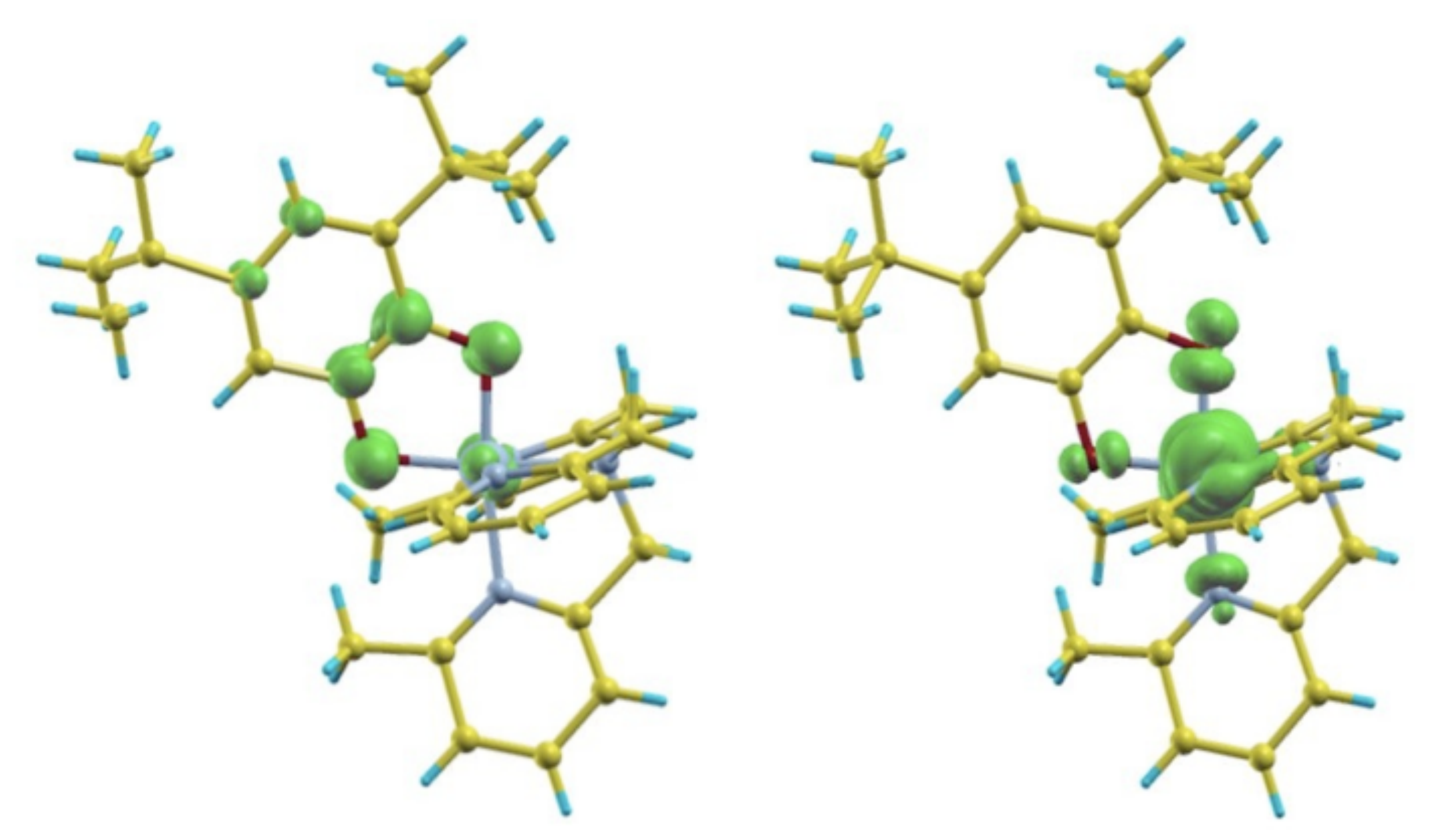}
\includegraphics[scale=0.48,clip=true]{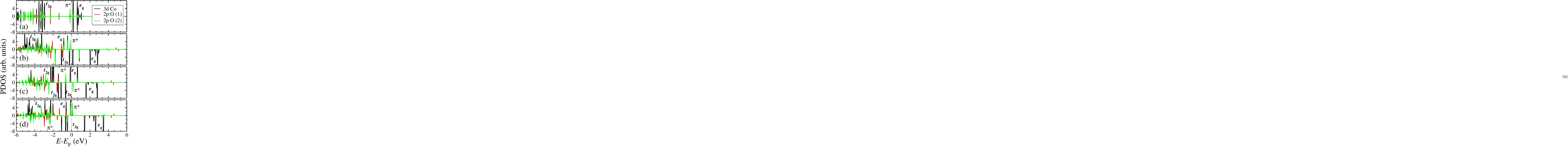}
\caption{Density of states projected over the $3d$ states of the Co and the $2p$ states of the two O atoms, as calculated with ASIC, for 
a) LS Co$^{3+}$-Cat b) HS Co$^{2+}$-SQ c) LS Co$^{2+}$-SQ and d) HS Co$^{2+}$-SQ with antiferromagnetic coupling between 
the magnetic moment of the Co and the one of the SQ. In the top panels we show the isovalue contour plot for the charge density 
(local DOS) for the HOMO (left) and LUMO (right) of the LS Co$^{3+}$-Cat.}
\label{Fig2}
\end{figure}
The ASIC calculated density of states (DOS) of the molecule is displayed in Fig. \ref{Fig2}. Although in general the 
details of the Kohn-Sham spectrum depend on the specific exchange and correlation functional used, in this case the relative 
position of the Co-$3d$ shell with the molecular orbitals of Cat (SQ) is returned consistently by both ASIC and B3LYP. Since, 
to our knowledge, there are no spectroscopic data for Co-dioxolene complexes to compare with our calculations, Fig. \ref{Fig2} 
should be interpreted as a schematic energy level diagram of the molecule. We notice that the self-interaction correction returns 
the expected weak hybridization between the Cat (SQ)-$\pi^*$ and the Co $3d$ states. The Co 3$d$ shell can be clearly recognized 
in the DOS and it is filled according to the formal oxidation. The symmetry of the highest occupied molecular orbital 
(HOMO) and the lowest unoccupied molecular orbital (LUMO) can be appreciated by looking at charge density isovalue contour 
plot. For instance (Fig.~\ref{Fig2}) for LS Co$^{3+}$-Cat the HOMO is the Cat-$\pi^*$ state while the LUMO corresponds to the 
Co $e_g$ states. This confirms the good level of the ASIC description.

\begin{figure}[ht]
\includegraphics[scale=0.35,clip=true]{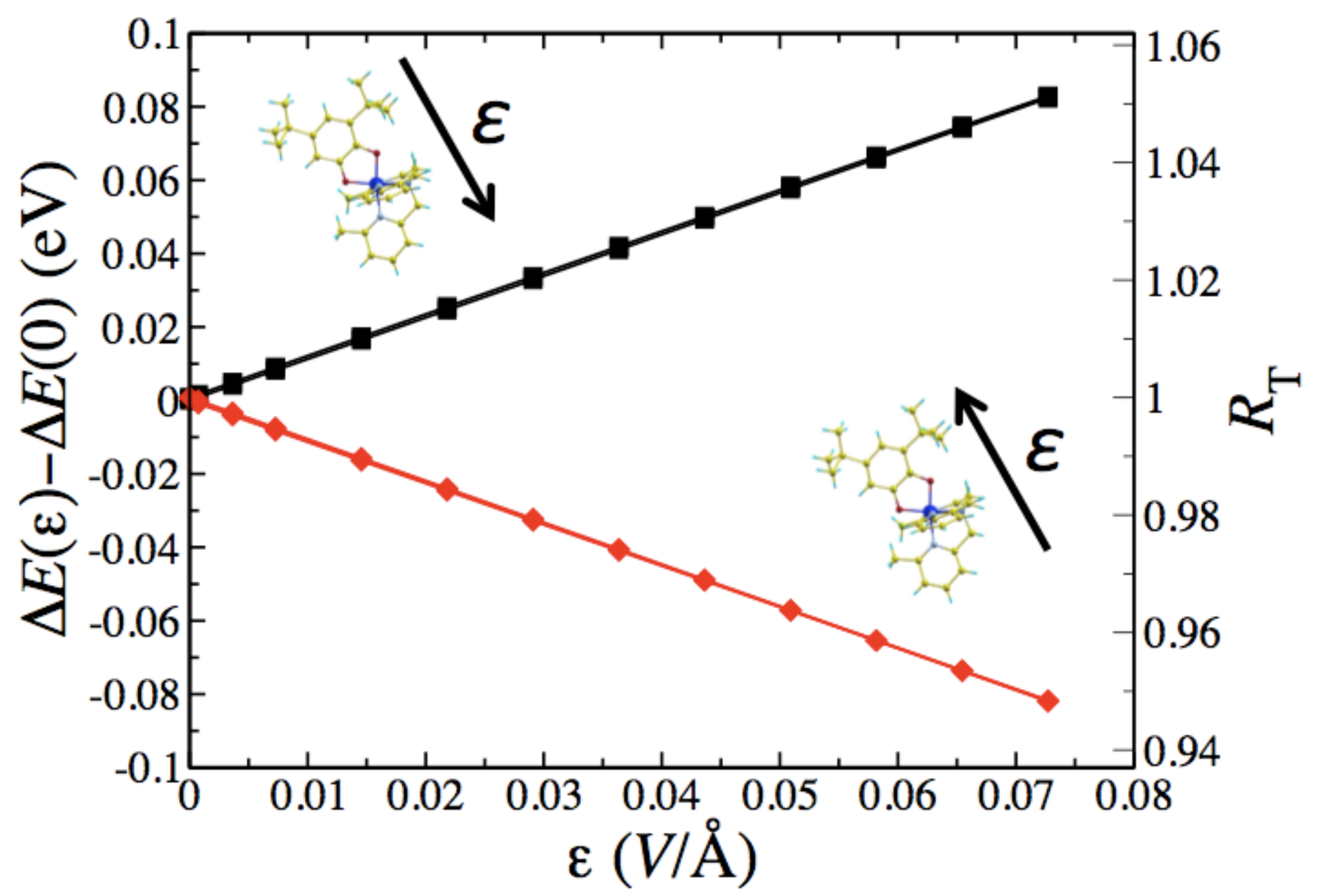}
\caption{Energy difference between HS Co$^{2+}$-SQ and LS Co$^{3+}$-Cat, $\Delta E$, as a function of the electric field $\mathcal{E}$ 
(the energies are calculated with the B3LYP functional). The black line (square symbols) corresponds to electric field pointing toward the 
quinone, while the red one (diamond symbols) to the electric field pointing toward the Co. The figure also report $R_\mathrm{T}$ for the 
same molecule on right-hand side $y$-axis.}
\label{Fig3}
\end{figure}

After having demonstrated that DFT provides a satisfactory description of the molecule electronic properties we now 
proceed to investigate the influence that an electric field, ${\mathcal E}$, has on the VTI. In particular we consider the 
situation where ${\mathcal E}$ is applied along the direction joining Co with Cat (SQ). In order to reduce the computational 
overheads we replace the N ligands with ammonia molecules. This substitution does not bring any relevant modification to 
the relative position of the Co-$3d$ and the Cat (SQ)-$\pi^*$ orbitals. However it has an effect on the energetics of the problem, 
since it decreases the energy difference between HS Co$^{2+}$-SQ and LS Co$^{3+}$-Cat to 1.591~eV. Figure \ref{Fig3} shows 
the difference, $\Delta E=E_\mathrm{HS Co^{2+}SQ}-E_\mathrm{LS Co^{3+}Cat}$, between the ground state energies of LS 
Co$^{3+}$-Cat and HS Co$^{2+}$-SQ ($\Delta E>0$ means $E_\mathrm{HS Co^{2+}SQ}>E_\mathrm{LS Co^{3+} Cat}$). 
Such a quantity is plotted as a function of the electric field and with respect to the $\mathcal E=0$ situation. The most important 
observation is that $\Delta E$ depends linearly on $\mathcal E$.

We recall here that the energy change, $\Delta E$, of a quantum mechanical system under the influence of an external electric 
field is governed by the Stark effect. This reads 
$\Delta E({\mathcal E})\propto\vec{p}\cdot\vec{\mathcal{E}}+\frac{1}{2}\sum_{ij}{{\mathcal E}_i}{\alpha}_{ij}{{\mathcal E}_j}$ 
where $\vec{p}$ is the permanent electrical dipole and ${\alpha}_{ij}$ the polarizability tensor. Thus the linear dependance found in 
Fig.~\ref{Fig3} indicates that the first order Stark effect dominates the molecule response. Our DFT calculations confirm such an
hypothesis and return a finite dipole moment at $\mathcal E=0$ ($|\vec{p}|$ is 16.9~Debye for LS Co$^{3+}$-Cat and 26.3~Debye for 
HS Co$^{2+}$-SQ). Interestingly the energy decreases when the electric field points from the Co to the dioxolene 
and increases when it is along the opposite direction. This means that the $\Delta E$ is reduced for the field direction facilitating the 
VTI electron transfer. A linear fit of $\Delta E({\mathcal E})$ gives us a slope of 1.12~eV/(V/\AA). We can now extrapolate a 
critical electric field ${\mathcal E}_\mathrm{C}=1.4$~V/\AA\ at which the states LS Co$^{3+}$-Cat and HS Co$^{2+}$-SQ become 
energetically degenerate. When ${\mathcal E}$ exceeds ${\mathcal E}_\mathrm{C}$ a VTI occurs leading to SC and 
HS Co$^{2+}$-SQ becomes the new ground state of the molecule. 

The calculated ${\mathcal E}_\mathrm{C}$ unfortunately is quite large, so that in reality one can not expect to observe an electric field induced VTI. 
However, it is important to note that such a large predicted ${\mathcal E}_\mathrm{C}$ is mainly the consequence of the large 
overestimation of $\Delta E({\mathcal E}=0)$, i.e. of the large DFT error over the energetics of the system in absence of an electric 
field. A much less sensitive quantity to DFT uncertainties is the slope of $\Delta E({\mathcal E})$, which depends only on the molecule 
electrical dipole and polarizability (accurately described in DFT). This will allow us to estimate with precision the relative change
in the VTI crossover temperature, $T_\mathrm{C}$, as a function of ${\mathcal E}$.

$T_\mathrm{C}$ is defined by the condition $\Delta G=0$, i.e. it corresponds to the situation where the Gibbs free energies of the
two spin configurations become identical. This yields to the equation
\begin{equation}\label{eq2}
T_\mathrm{C}=\frac{\Delta H}{\Delta S}=\frac{\Delta E}{\Delta S}\:,
\end{equation}
where we have set the pressure to zero since we are considering the molecule ``in vacuum'', and we have approximated the enthalpy 
at $T_\mathrm{C}$ with the zero-temperature total electronic energy. In writing the equation (\ref{eq2}) we have also assumed that the 
entropy variation, $\Delta S$,  does not depend on the electric field, i.e $\Delta S(\mathcal E)\approx\Delta S(0)$. This is consistent with 
our computational strategy of not relaxing the atomic coordinates. We have verified in a few cases that the atomic relaxation under 
bias is much smaller then that due to the VTI, so that our assumption is well grounded in our own calculations. 

Then, some algebra shows that the ratio between the critical temperature calculated in an electric field, $T_\mathrm{C}(\mathcal E)$, with 
that in zero field, $T_\mathrm{C}(0)$, simply writes
\begin{equation}
R_\mathrm{T}=\frac{T_\mathrm{C}(\mathcal E)}{T_\mathrm{C}(0)}=\frac{\Delta E(\mathcal E)}{\Delta E(0)}\label{T_c_ratio}\;.
\end{equation}
Importantly $R_\mathrm{T}$ is expressed in terms of DFT total energies only, and in Fig. \ref{Fig3} we show that changes of $T_\mathrm{C}$ 
of the order of 5\% can be obtained already at the much more modest field of 0.07~eV/\AA. 

The uncertainty over such a value nevertheless remains quite large since the calculation still involves the precise determination 
of $\Delta E(0)$. A better estimate can be obtained by eliminating $\Delta E(0)$ completely and just use information about the slope of 
$\Delta E({\mathcal E})$. This can be obtained by re-writing the equation (\ref{eq2}) as 
\begin{equation}
\left[T_\mathrm{C}(0)-T_\mathrm{C}(\mathcal E)\right] \Delta S(0)= \Delta E(0)- \Delta E(\mathcal E)\:.
\end{equation}
We now take the experimental estimate\cite{Beni} for the entropy variation $\Delta S(0)\approx 0.52$~meV/K, and
conclude that achievable electric fields can indeed induce large variations in $T_\mathrm{C}$. In fact an electric field 
of 0.1~V/nm, typical of Stark spectroscopy \cite{SS}, produces already a change in $T_\mathrm{C}$ of about 21~K, and a field of
0.5~V/nm, obtainable in scanning tunnel microscopy experiments can drift $T_\mathrm{C}$ by as much as 100~K. 

In conclusion we have demonstrated that the critical temperature for the VTI in Co-dioxolene complexes can be drastically modified by
a static electric field. Intriguingly, this represents a physical way to effectively modulate the redox potential of the metal acceptor, alternative to 
the chemical strategy of changing the radical groups of the molecule \cite{Beni}. Such an external control of the magnetism of valence 
tautomeric compounds may open new avenues to the growing field of molecular spintronics. 

We would like to thank N. Baadji, G. Poneti, R. Sessoli and A. Dei for useful discussions. This work is sponsored by 
Science Foundation of Ireland (07/RFP/MASF238). Computational resources have been provided by TCHPC.

\end{document}